\documentclass[times,trackchanges]{aastex631} 

\submitjournal{ApJ}

\usepackage{savesym}
 
\restoresymbol{SIX}{tablenum}
\usepackage{booktabs}
\usepackage{multirow}
\usepackage{physics}
\usepackage{textcomp}
\usepackage{hyperref}
\usepackage[capitalize]{cleveref}
\usepackage{amssymb}
\usepackage{xcolor}
\usepackage{graphicx}  
\usepackage{longtable} %
\usepackage{makecell} %
\usepackage{tabularx}

\makeatletter
\usepackage{etoolbox}
\patchcmd\H@refstepcounter{\protected@edef}{\protected@xdef}{}{}
\makeatother

\graphicspath{{./}{assets/}}

\begin{document}

\title{Wave Activity at MHD–ion Scales Associated with Switchbacks}
\shorttitle{Wave activity associated with Switchbacks}
\shortauthors{Choi et al.}

\author[0000-0003-2054-6011]{Kyung-Eun Choi}
\affiliation{Space Sciences Laboratory, University of California Berkeley: Berkeley, CA, USA}
\author[0000-0001-6427-1596]{Oleksiy V. Agapitov}
\affiliation{Space Sciences Laboratory, University of California Berkeley: Berkeley, CA, USA}
\affiliation{Astronomy and Space Physics Department, National Taras Shevchenko University of Kyiv, Kyiv, Ukraine}

\author[0000-0002-2011-8140]{Forrest Mozer}
\affiliation{Space Sciences Laboratory, University of California Berkeley: Berkeley, CA, USA}

\author[0009-0000-8639-188X]{Seung-Ju Yang}
\affiliation{Department of Astronomy and Space Science, Chungbuk National University, Cheongju, Republic of Korea}
\author[0000-0001-9994-7277]{Dae-Young Lee}
\affiliation{Department of Astronomy and Space Science, Chungbuk National University, Cheongju, Republic of Korea}

\author[0000-0001-9254-3149]{Richard D. Sydora}
\affiliation{Department of Physics, University of Alberta, Edmonton, Canada}

\author[0000-0001-6016-7548]{Lucas Colomban}
\affiliation{Space Sciences Laboratory, University of California Berkeley: Berkeley, CA, USA}
\affiliation{Atomic Energy Commission, Paris, France}

\author[0000-0001-9448-0030]{Liudmyla Kozak}
\affiliation{Astronomy and Space Physics Department, National Taras Shevchenko University of Kyiv, Kyiv, Ukraine}

\author[0000-0003-2981-0544]{Mingzhe Liu}
\affiliation{Space Sciences Laboratory, University of California Berkeley: Berkeley, CA, USA}
\author[0000-0002-1573-7457]{Marc Pulupa}
\affiliation{Space Sciences Laboratory, University of California Berkeley: Berkeley, CA, USA}
\author[0000-0002-9954-4707]{Jia Huang}
\affiliation{Space Sciences Laboratory, University of California Berkeley: Berkeley, CA, USA}

\author[0000-0002-5121-600X]{Shaosui Xu}
\affiliation{Space Sciences Laboratory, University of California Berkeley: Berkeley, CA, USA}

\correspondingauthor{Kyung-Eun Choi}
\email{kechoi@berkeley.edu}


\begin{abstract}

Magnetic switchbacks (SB) -- the localized magnetic structures with magnetic field direction inclined at an angle $\theta$ relative to the background $B_0$ -- in the young solar wind have been associated with enhanced ion-scale wave activity and local plasma heating. 
It remains debated whether the apparent wave-power increase is intrinsic or mainly caused by sampling geometry. 
In this work, we analyze magnetic and electric field fluctuations measured by Parker Solar Probe, focusing on the 0.1--3~\(f_{cp}\) frequency band that spans the transition from the MHD inertial range to ion-kinetic scales. 
By decomposing magnetic fluctuations into field-aligned and transverse components and comparing SB and non-SB intervals at the same local magnetic field angle, we test whether SBs sample an anisotropic cascade from different viewing angles or host intrinsically amplified wave activity. 
We find that the transverse magnetic power $\delta B_{\perp}$ is systematically enhanced inside switchbacks across a wide range of magnetic field rotation angles $\theta$.
The enhancement persists even at small and intermediate deflections, where geometric projection alone predicts weak power, indicating an intrinsic origin beyond sampling geometry.
The inertial-range spectral indices also remain similar between SB and non-SB intervals despite the enhanced wave power inside SBs, suggesting that the underlying turbulence cascade is largely preserved.
This excess $\delta B_{\perp}$ coincides with elevated proton temperatures and enhanced electric-field fluctuations, supporting the interpretation that SBs act as localized sites of cross-scale energy transfer and ion-scale dissipation in the near-Sun solar wind.
\end{abstract}


\section{Introduction}\label{sec:intro}    %

Magnetic switchbacks (SBs) are mesoscale structures in the solar wind, characterized by sharp, large-angle rotations of the magnetic field direction relative to the surrounding Parker-spiral field, while its magnitude remains nearly constant and the radial flow speed is enhanced \citep{bale2019, kasper2019, Krasnoselskikh2020, agapitov2023}.
Such large deflections had already been reported at heliocentric distances between roughly 0.3 and 1 au by the Helios missions \citep[e.g.,][]{horbury+2018}, and at 1 au and beyond by spacecraft such as ACE and Ulysses, where they appear as intermittent radial-field reversals superposed on the Parker spiral \citep[e.g.,][]{kahler_lin_1994, crooker+2004, owens+2013}.

Parker Solar Probe (PSP) has shown that in the young solar wind, inside about 50~\(R_S\), SB-like rotations become much more frequent, indicating that folded magnetic-field structures are a common ingredient of the near-Sun solar wind \citep[e.g.,][]{Badman2026, wyper2026magnetic}. 
Statistical PSP studies further demonstrate that these SBs occur over a broad range of heliocentric distances in the inner heliosphere and tend to organize into patches or sequences, sometimes in association with locally generated waves \citep{Mozer2020, agapitov_sunward-propagating_2020, Krasnoselskikh2020,choi2024whistler, colomban_quantifying_2024, karbashewski2023whistler, froment2023whistler}, meso-scale structures such as small-scale magnetic flux ropes and blobs \citep{choi_series_2024, choi2025_switchbacks_SMFRs}, and fast-wind streams \citep{horbury+2020,fargette_preferential_2022, zank+2020}, suggesting that they play a non-negligible role in the large-scale dynamics and energetics of the corona–solar-wind system.

Recent PSP observations show that ion-scale waves, including ion cyclotron waves (ICWs) and kinetic Alfvén waves (KAWs), are commonly observed within SBs and along their sharp boundaries \citep{Mozer2020, bowen_ion-scale_2020, verniero_parker_2020}.
These ion-scale waves are likely driven by a combination of strong velocity shear, sharp field rotations, and kinetic free energy in the ion distributions. SB boundaries, in particular, have been shown to behave as shear layers that can support KHI-driven surface waves at MHD scales \citep{choi2025_surface}, while reconnection-associated magnetic structures have also been identified in the near-Sun solar wind \citep{Lee2026}, highlighting the importance of velocity shear in structuring the wave activity around SBs.
At larger scales, SBs themselves are embedded in an Alfvénic turbulent background, and their internal structure is often consistent with large-amplitude MHD Alfvén waves or flux-tube–like perturbations \citep{horbury+2020,squire_-situ_2020,mallet2021}, and the ion-scale wave activity they host can be viewed as a possible continuation of the MHD cascade into the kinetic range.


Ion-scale waves associated with SBs are expected to contribute to the dissipation of turbulent energy \citep{bale2019, agapitov_sunward-propagating_2020, cattell_modeling_2021}.
Through local ion heating and electron scattering, they provide a channel for converting fluctuation energy into particle thermal energy.
In addition, proton temperature anisotropies and nonthermal ion distributions observed near SBs and ion-scale waves point to ongoing kinetic regulation and resonant heating in these regions \citep{woolley2021, bale2009}.
Moreover, PSP measurements indicate that SBs can carry enhanced radial Poynting flux and kinetic energy flux compared to the surrounding solar wind \citep{Mozer2020, Woolley2020}, reinforcing the view that SBs act as efficient channels for transporting electromagnetic and flow energy away from the Sun.

The coexistence of SBs with ion-scale fluctuations has been interpreted as evidence that SBs may serve as active mediators between large-scale MHD turbulence and ion-kinetic processes \citep{bowen2020constraining, squire_-situ_2020, mallet2021}.
In this picture, SBs are not only large-scale Alfvénic structures but also important sites for energy transfer across scales. 

Some studies have also shown that including SB intervals in the analysis leads to an apparent enhancement of magnetic fluctuation power and changes in the turbulent spectral indices, suggesting that part of the excess power associated with SBs may be geometric or selection-driven \citep{dudok_de_wit2020}.
More recently, it has been argued that the apparent enhancement of wave power across the inertial and kinetic frequency ranges inside SBs may partly arise from geometric projection effects \citep{tatum2024}. \cite{tatum2024} examined how the magnetic fluctuation power within SBs depends on the angle between the magnetic field and the solar wind flow, and showed that the resulting geometric sampling effects largely account for the observed increase in wave power, implying that SBs do not strongly modify the local turbulent cascade via wave–particle interactions.

In the present study, we extend this perspective by analyzing fluctuations in a field-aligned coordinate system and by incorporating plasma-frame electric-field measurements. These measurements enable a more direct comparison of the parallel and perpendicular components of wave activity inside and outside SBs.
We focus on fluctuations in the frequency range of 0.1--3~\(f_{cp}\), which lies in the transition from the MHD inertial range to ion-kinetic scales where ion-scale effects begin to play an important role \citep{chen2014,franci2015,bowen2020constraining}.
In this regime, Alfvénic fluctuations are expected to be modified by kinetic processes, giving rise to modes such as ion cyclotron and kinetic Alfvén waves \citep{fisk_Gloeckler_2006, jian2009, he2011, duan2021}.
By examining how the power in this band evolves inside and outside SBs as a function of heliocentric distance, we aim to assess whether these structures mediate the transfer of energy across scales and contribute to dissipation in the near-Sun solar wind \citep{squire_-situ_2020,bowen2020constraining, mallet2021, bandyopadhyay2020}.
In particular, by decomposing fluctuations into field-aligned components and explicitly controlling for magnetic-field geometry, we test whether the enhanced wave activity reported inside SBs can be fully explained by geometric sampling or instead reflects intrinsic amplification of ion-scale fluctuations within these structures.

The paper is organized as follows. In Section 2, we describe the PSP data sets and our field-aligned and plasma-frame analysis methods. 
In Section 3, we present case studies and statistical results for wave power inside and outside SBs. 
In Section 4, we discuss the implications for MHD ion-scale coupling and dissipation in the near-Sun solar wind, and summarize our main conclusions.

\section{Data and Methods} 
\subsection{Parker Solar Probe Data}
In this study, we use magnetic field, electric field, and plasma measurements from the Parker Solar Probe (PSP). 
The Magnetic and electric field observations are provided by the Fluxgate Magnetometer (MAG) and the Electric Field Instrument \citep[EFI;][]{Mozer2020} that are part of the FIELDS suite \citep{bale+2016}. 
The plasma parameters are obtained from the Solar Wind Electrons Alphas and Protons investigation \citep[SWEAP;][]{kasper+2016}, which consists of the Solar Probe Cup \citep[SPC;][]{case+2020} and Solar Probe Analyzers sensors for ions and electrons \citep[SPAN-i and SPAN-e;][]{livi+2022,whittlesey+2020}. Electron number density derived from Quasi-Thermal-Noise (QTN) spectroscopy is also used herein \citep{2020ApJSMoncuquet}, which has served as a critical calibration standard in numerous previous scientific studies \citep[e.g.,][]{2021PhRvLKasper,2021LiuAA,2023LiuAA,2021ApJZhao}.

\subsection{Switchback Catalog and $\theta$ Definition}
The local field angle ($\theta$) is determined as the angular deviation of the magnetic field vector from the moving-averaged magnetic field with a 2.4-hour window. 
Averaging is applied only to the ambient solar wind interval, excluding SB events. 
We identify SB intervals for Encounter 1 (E01) using existing event lists \cite{huang2023} and \cite{agapitov2023}, and construct a new catalog for Encounters 17-21 (E17--21) following the criteria of \cite{huang2023}.

We use E01 of PSP as a baseline interval to characterize the local field-angle dependence of wave power at 35--50~\(R_S\). To examine how these properties vary with heliocentric distance while avoiding bias from a single encounter, we also analyze E17--21, which span between about 11--50~\(R_S\) between PSP’s sixth and seventh Venus flybys and repeatedly sample the same radial range. 
This configuration provides a large number of SB intervals for statistical analysis of the wave-geometry relationship, and we present results separately for the 35--50~$R_S$ and 11--35~$R_S$ ranges.

We develop an automatic detection algorithm following the procedure described by \cite{huang2023}, with adaptations for our event-catalog construction: (i) identify deflections of the local magnetic field relative to the background magnetic field ($B_0$), where $B_0$ is determined from the ambient solar wind; (ii) define intervals based on the fractional deviation from $B_0$: Quiet (85\% of $B_0$), Transition (85-15\%), and Spike ($<$15\%), (iii) confirm magnetic polarity using suprathermal electron pitch-angle distributions (PADs) through the asymmetry parameter $\epsilon = ({J_{\parallel}-J_{\rm anti}})/
({J_{\parallel}+J_{\rm anti}})$, where $J_{\parallel}$ and $J_{\rm anti}$ denote electron fluxes at 0$^\circ$-90$^\circ$ PAD and 90$^\circ$-180$^\circ$ PAD, respectively, classified as positive ($\epsilon >$ 0.6) or negative ($\epsilon<$ -0.6); and (iv) if the Spike period persists for more than 3 seconds, we classify it as an SB event.

\subsection{Wave Power Analysis}
For the spectral analysis, magnetic and electric field data are processed using 3-ms and 6-ms cadence measurements, respectively, and power spectra are evaluated over 0.1--3~\(f_{cp}\), corresponding to frequencies up to a few tens of Hz near 0.05 au (11.4~\(R_S\)). 
In the frequency range 0.1--3~\(f_{cp}\), it spans the transition from the MHD inertial range into ion-kinetic scales. The lower end ($\sim0.1f_{cp}$) corresponds to the break region where the magnetic power spectrum steepens. 
Proton-scale effects start to modify the MHD cascade, whereas the upper end ($\sim$1-3~\(f_{cp}\)) is where ion cyclotron and kinetic Alfvén waves, as well as anisotropy-driven ion-scale instabilities, are most commonly observed. 
By isolating this band, we specifically target the range in which large-scale Alfvénic energy associated with SBs can be transferred into ion-scale wave activity, rather than being dominated by lower-frequency MHD structures or higher-frequency electron-scale fluctuations. 
Because Doppler shifting affects the spacecraft-frame frequencies, the 0.1--3~\(f_{cp}\) interval receives power from a range of wave vectors and propagation directions near proton scales, and we therefore treat it as a transition band rather than a single, well-defined spatial scale.
The integrated power is used as a measure of the ion-scale wave activity.

\subsection{Spectral Analysis}

The spectral analysis was performed using fixed 1-s  analysis windows. 
For each window, the magnetic field spectral index was determined from the power spectrum over the 0.01--1 Hz range and assigned a representative field angle, $\theta$, based on the average local magnetic field direction. 
The intervals were subsequently classified as SB or non-SB, and the resulting spectral indices were used to construct the statistical distributions presented in Section~\ref{sec:turbulence}.

\section{Observations of Wave Activity associated with Switchbacks}
\subsection{Switchback events and wave activity signatures}  
\begin{figure*}
    \centering
    \includegraphics[width=\textwidth]{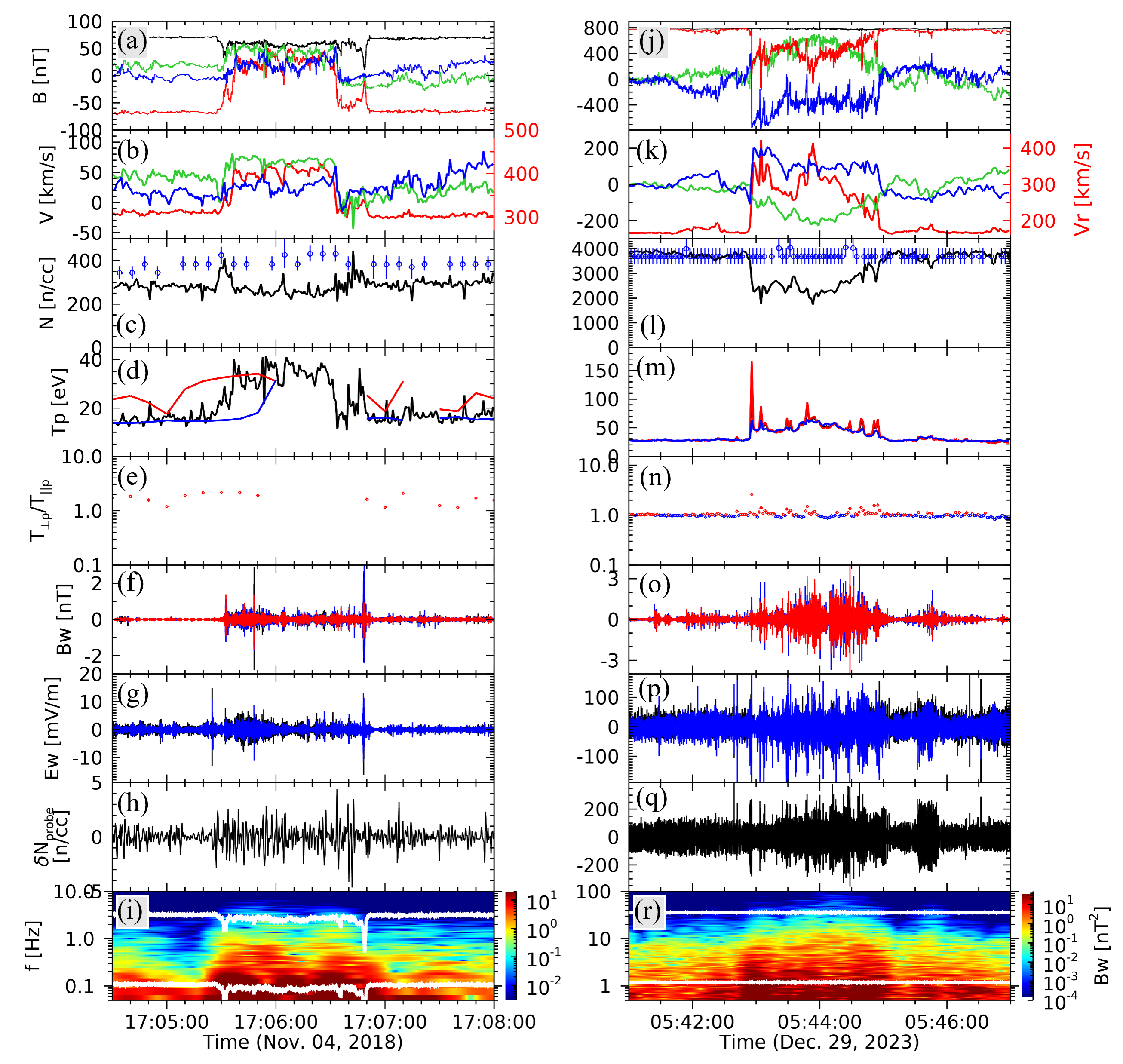}
    \caption{Representative SB intervals observed by PSP at two heliocentric distances. The left column shows an event at 37.2~\(R_S\)(Nov. 4, 2018; E01), and the right column shows an event at 11.8~\(R_S\)(Dec. 29, 2023; E18). 
    Panels (a,j) show the magnetic field components in RTN coordinates (black: magnitude, red: R, green: T, and blue: N), and panels (b,k) show the corresponding plasma bulk velocity components (the radial component is indicated on the right axis). Panels (c,l) show the proton density (black) together with electron density from quasi-thermal noise estimates (blue). Panels (d,m) display the proton temperature (black), perpendicular temperature $T_{\perp}$ (red), and parallel temperature $T_{\parallel}$ (blue) relative to the magnetic field. Panels (e,n) show the proton temperature anisotropy ratio ($T_{\perp}/T_{\parallel}$), with red (blue) indicating values greater (less) than unity. respectively. Panels (f,o) and (g,p) present the magnetic ($B_w$) and electric ($E_w$) field fluctuations, respectively, and panels (h,q) show the spacecraft potential fluctuations. Panels (i,r) show the magnetic-field power spectrograms. The two white curves mark 0.1~\(f_{cp}\) and 3~\(f_{cp}\), where \(f_{cp}\) is the local proton cyclotron frequency. Enhanced power in this band is observed during the SB intervals.}
    \label{fig:example}
\end{figure*}
\Cref{fig:example} presents two representative SB intervals observed by PSP at different heliocentric distances: 37.2~\(R_S\) during E01 (left column) and 11.8~\(R_S\) during E18 (right column), where two successive SBs occur. In these examples, SBs are characterized by rapid changes in the magnetic field direction, accompanied by variations in the plasma bulk velocity as shown in Figures \ref{fig:example}(a,b,i, and j). The magnetic field deflections are primarily rotational rather than compressional, as indicated by the relatively stable magnetic field magnitude. The velocity vector exhibits correlated variations consistent with Alfvénic behavior. These features confirm the Alfvénic nature of the switchbacks at both radial distances.

Figures \ref{fig:example}(c,l) show the proton density ($N_p$) and the electron density estimated from quasi-thermal noise ($N_e$), both of which exhibit modest variations across the SB intervals. A reduction of Np is visible inside the SB core, while the QTN-derived electron density shows a comparable but not identical trend. Previous statistical studies, primarily covering encounters up to E10, reported that the interior and exterior densities of SBs are generally similar, based on electron densities \citep{jagarlamudi_occurrence_2023} and proton densities \citep{larosa_switchbacks_2021, martinovic2021}. Differences between $N_p$ and $N_e$ may partly reflect composition and measurement differences, including the alpha-particle contribution under quasi-neutrality.
The proton temperature behavior associated with SBs (Figures \ref{fig:example}(d,m)) is addressed in detail in the following subsection.

Notably, enhanced wave-like perturbations are observed in both the magnetic ($B_w$) and electric-field ($E_w$) fluctuations during the SB intervals. Figures \ref{fig:example}(i,r) show the corresponding magnetic field power spectra, where enhanced power appears during the SB intervals. The two white curves indicate the local proton cyclotron frequency, 0.1~\(f_{cp}\) and 3~\(f_{cp}\), respectively.
Figures \ref{fig:example}(h,q) also display perturbations in the high-frequency electron density estimated from spacecraft surface charging, which provides an independent proxy for rapid plasma/electric variability, based on internal FIELDS calibration analyses \citep[e.g.,][AGU Fall Meeting]{Liu2025AGU}. We use this quantity here only as a contextual indicator of high-frequency activity.

\subsection{Temperature Behavior Inside Magnetic Switchbacks}  
PSP observations consistently reveal that the internal environment of magnetic SBs is generally characterized by enhanced proton temperatures compared to the ambient solar wind \citep{Luo2023, Badman2026,Krasnoselskikh2020, Mozer2020, Rasca2021, huang2023}. Specific SB events exhibit distinct temperature spikes that can exceed 50\% or even quadruple the external ion temperature, particularly at boundaries with sharp magnetic field dropouts \citep{Farrell2020, Krasnoselskikh2020}. This heating varies significantly in its anisotropy; while some statistical studies identify nonadiabatic heating in both parallel and perpendicular directions \citep{Luo2023}, others associate SB patches primarily with enhanced parallel temperatures consistent with reconnection formation mechanisms \citep{Woodham2021}, or conversely, attribute temperature increases to stronger perpendicular heating driven by turbulence dissipation \citep{Shi2022}. \cite{Woolley2020} reported that the proton core parallel temperature can remain unchanged, following a rigid velocity space rotation, or that apparent temperature modulations are geometric effects revealing a dominant perpendicular anisotropy at oblique angles \citep{Raouafi2023}. 

However, recent analyses suggest that deflections often correlate with increased parallel temperatures leading to isotropy \citep{Laker2024}, and that these magnetically isolated structures, potentially formed by flux rope merging, exhibit enhanced thermal velocities and Alfvénicity \citep{Agapitov2022}. Ultimately, these temperature enhancements, which may result from scattered bulk-velocity protons or Poynting fluxes \citep{Mozer2020}, and involve competing heating mechanisms where alpha particles and protons display opposite temperature trends between spike and transition regions \citep{huang2023}.

Figures \ref{fig:example}(d,e,m,n) show elevated proton temperatures within representative SB intervals relative to the surrounding solar wind, with pronounced enhancements accompanied by sharp magnetic deflections. 
The temperature anisotropy ratio ($T_{\perp}/T_{\parallel}$) also exhibits noticeable variations within these intervals.
Critically, these temperature enhancements coincide with intervals of enhanced magnetic and electric fluctuations and wave power increases described above, suggesting a possible link between intensified wave activity and local plasma heating.
This motivates examining whether the wave-power enhancement is primarily controlled by the sampling geometry or reflects intrinsic amplification within SBs.
In the following sections, we explicitly relate these temperature enhancements to the band-integrated wave power in the 0.1--3~$f_{cp}$ range, in order to test whether local proton heating inside SBs systematically tracks intrinsically amplified ion-scale fluctuations rather than arising solely from geometric sampling effects.

\subsection{Wave power dependence on the magnetic field geometry} 

This section analyzes the dependence of the decomposed wave power on the local field angle ($\theta$) using data from E01 and E17--21, which provides a representative near-Sun example with clear SB activity and well-defined angular variations. The first part characterizes how the parallel and perpendicular power components vary with $\theta$ (defined in Section 2) to establish the geometric baseline. The second part compares the power inside and outside SBs for similar $\theta$ values to determine whether the observed enhancement results purely from geometric sampling or also involves intrinsic wave activity.


\begin{figure*}
    \centering
    \includegraphics[width=\textwidth]{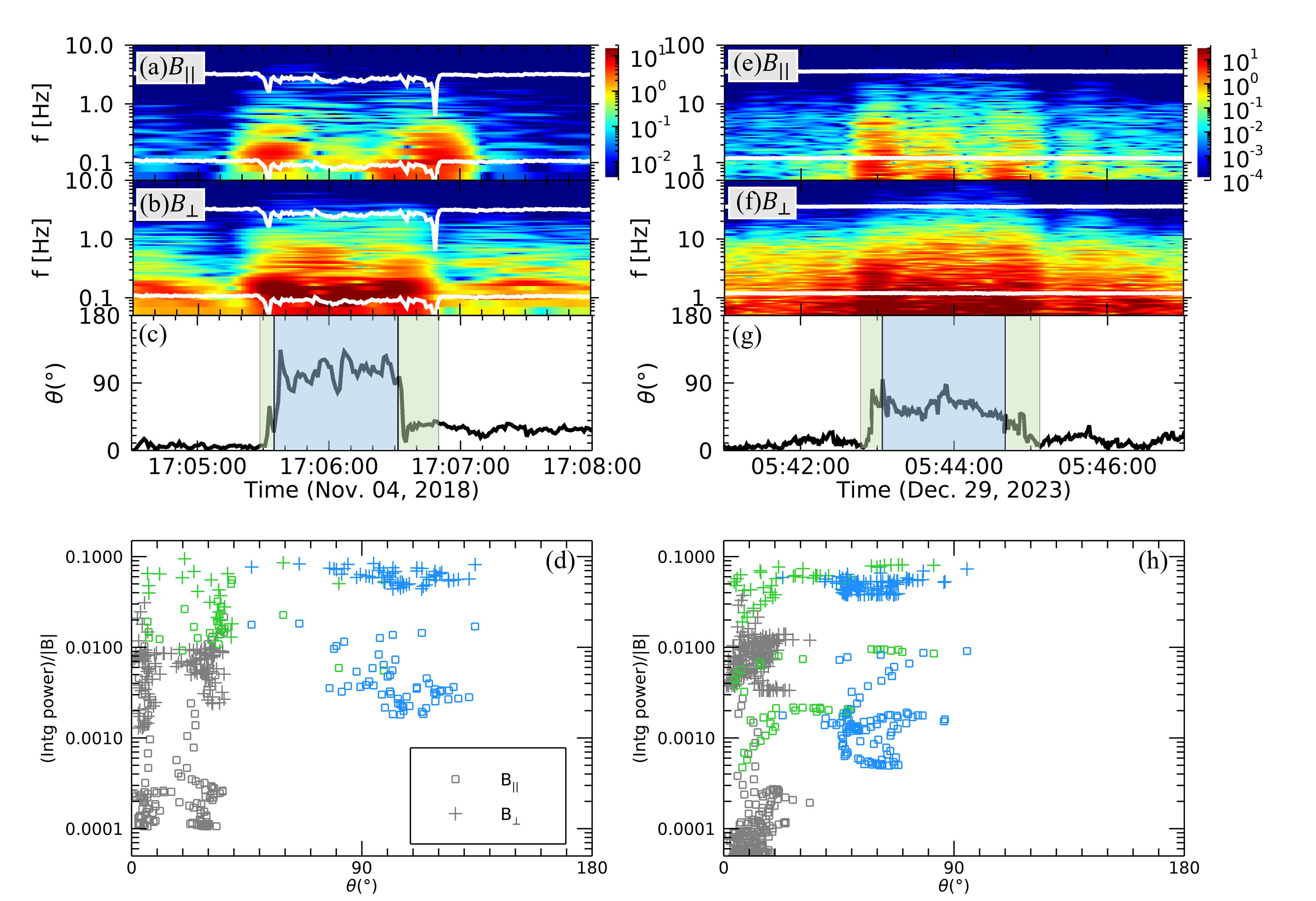}
    \caption{Field-aligned wave activity associated with representative SB intervals shown in \Cref{fig:example}. The left column shows an event at 37.2~\(R_S\)(Nov. 4, 2018; E01), and the right column shows an event at 11.8~\(R_S\)(Dec. 29, 2023; E18). Panels (a,e) show the magnetic-fluctuation power spectrograms of the field-parallel component ($B_{\parallel}$), and panels (b,f) show those of the field-perpendicular component ($B_{\perp}$). Panels (c,g) display the corresponding local field angle ($\theta$), with shaded regions indicating SB intervals (blue: SB core; green: boundary). Panels (d,h) show the normalized band-integrated power over 0.1–3~\(f_{cp}\) (marked by the two white curves in spectrograms) as a function of $\theta$, where open squares denote $B_{\parallel}$ and crosses denote $B_{\perp}$. The power is normalized by the local magnetic-field magnitude. Symbols are colored for the SB core (blue) and boundary (green) intervals, while gray indicates the ambient solar wind.}
    \label{fig:example_FAC}
\end{figure*}

To investigate whether the enhanced wave activity is controlled by magnetic-field geometry, we first examine the temporal variation of the local field angle and the corresponding integrated power during representative SB intervals. \Cref{fig:example_FAC} illustrates how the wave power varies as a function of $\theta$, allowing us to establish a geometric baseline before separating SB and non-SB intervals. 
The spectrograms of the field-parallel component (Figures \ref{fig:example_FAC}(a,e)) exhibit a localized enhancement near the SB boundaries (green shaded intervals), although its overall level remains weaker than the field-perpendicular component. In contrast, the field-perpendicular component (Figures \ref{fig:example_FAC}(b,f)) shows a pronounced intensification within the SB core intervals (blue shaded intervals), indicating that the strongest wave activity is concentrated during the largest field deflections. 
Consistently, the normalized band-integrated power integrated over 0.1--3~\(f_{cp}\) captures these behaviors in Figures \ref{fig:example_FAC}(d,h). The $B_{\perp}$ power is elevated during the SB core, whereas the $B_{\parallel}$ power preferentially increases near the boundary sub-intervals, potentially providing favorable conditions for wave-particle interactions through enhanced magnetic inhomogeneity \citep{vo2024}. 


\begin{figure*}
    \centering
    \includegraphics[width=\textwidth]{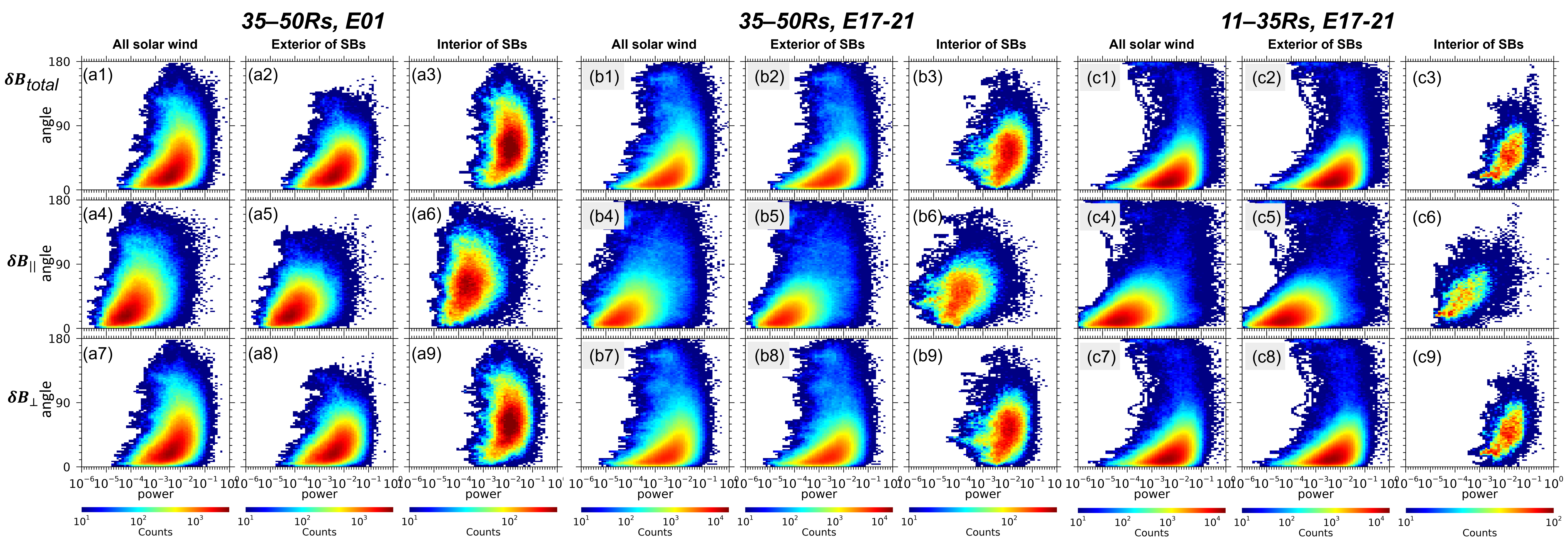}
    \caption{
    Statistical distribution of band-integrated magnetic wave power as a function of the local field angle $\theta$ for E01 (35--50~\(R_S\)) and encounters E17--21 at two heliocentric distance ranges (35--50~\(R_S\) and 11--35~\(R_S\)). The three column groups correspond to E01 (panels a1-a9), E17--21 at 35--50~\(R_S\) (panels b1-b9), and E17--21 at 11--35~\(R_S\) (panels c1-c9); within each group, the three columns show all solar-wind intervals (left), exterior of SBs (middle), and interior of SBs (right). The top row (a1-c3) presents the total MHD-band wave power, the middle row (a4-c6) shows the compressional component ($\delta B_{\parallel}$), and the bottom row (a7-c9) shows the transverse component ($\delta B_{\perp}$). The wave power is integrated over the 0.1--3~$f_{cp}$ range and normalized by the local magnetic-field magnitude $|\textbf{B}|$. Colors indicate occurrence counts on a logarithmic scale.}
    \label{fig:Bpower}
\end{figure*}

Figures \ref{fig:Bpower}(a1-a9) show the distribution of magnetic wave power on the local field angle $\theta$ for the E01 interval between 35 and 50~\(R_S\). 
In the first column (all solar wind intervals), the total MHD-scale wave power (Figure \ref{fig:Bpower}(a1)) increases with the field angle $\theta$. 
However, a clearer distinction emerges when the power is decomposed into compressional and transverse components.
The compressional wave power (Figure \ref{fig:Bpower}(a4)) exhibits relatively weak angular dependence and remains low even at large field angles.
In contrast, the transverse component (Figure \ref{fig:Bpower}(a7)) shows a pronounced enhancement for angles between about 80$^\circ$ and 150$^\circ$, indicating that wave activity is dominated by fluctuations perpendicular to the local magnetic field.


We next compare the exterior and interior SB intervals (second and third columns of \Cref{fig:Bpower}). In the first row (Figures \ref{fig:Bpower}(a2, a3)), the SB interior spans a broad range of angles but is most concentrated between 40$^\circ$ and 90$^\circ$. For comparable $\theta$ values, the interior intervals tend to exhibit systematically higher normalized power than the exterior intervals. While the overall angular trend remains, inside SBs, the dependence on $\theta$ appears weaker, with a reduced spread of power across the angular range compared to the exterior intervals.

We then examine whether this enhancement is associated with specific fluctuation components (middle and bottom rows). The compressional component, $\delta B_{\parallel}$ (Figures \ref{fig:Bpower}(a5, a6)), is typically weaker than the transversal component $\delta B_\perp$ (Figures \ref{fig:Bpower}(a8, a9)). Nevertheless, the $\delta B_{\parallel}$ distributions in the SB interior (\Cref{fig:Bpower} (a6)) occasionally extend to power levels comparable to $\delta B_\perp$ within the same $\theta$ bins, particularly at low to intermediate angles (e.g., $\theta < 60^\circ$).

As shown in \Cref{fig:Bpower} (a9), in contrast, $\delta B_\perp$ exhibits elevated power inside SBs across a broad angular range, with a noticeably weaker dependence on $\theta$ than outside SBs. This behavior suggests that the enhanced power cannot be explained solely by sampling geometry and is primarily associated with intensified perpendicular fluctuations. 
To examine whether these properties persist at smaller heliocentric distances, we extend the same analysis to E17--21.

Applying the same analysis to E17--21 (Figures \ref{fig:Bpower} (b-c)) shows that overall angular trends persist at smaller heliocentric distances.
For both 35--50~\(R_S\) and 11--35~\(R_S\), the total MHD-band power r is dominated by the transverse component, with $\delta B_\perp$ generally increasing with $\theta$, while $\delta B_{\parallel}$ remains comparatively weak. 
However, unlike the surrounding solar wind, the SB interiors exhibit a much weaker dependence on $\theta$, while maintaining systematically elevated  $\delta B_\perp$ power across a broad angular range. 
This behavior persists from E01  down to $\sim$11~\(R_S\), indicating that enhanced perpendicular fluctuations within SBs are a robust feature of the near-Sun solar wind.

\subsection{Wave power variation with heliocentric distances}

\begin{figure*}
    \centering
    \includegraphics[width=\textwidth]{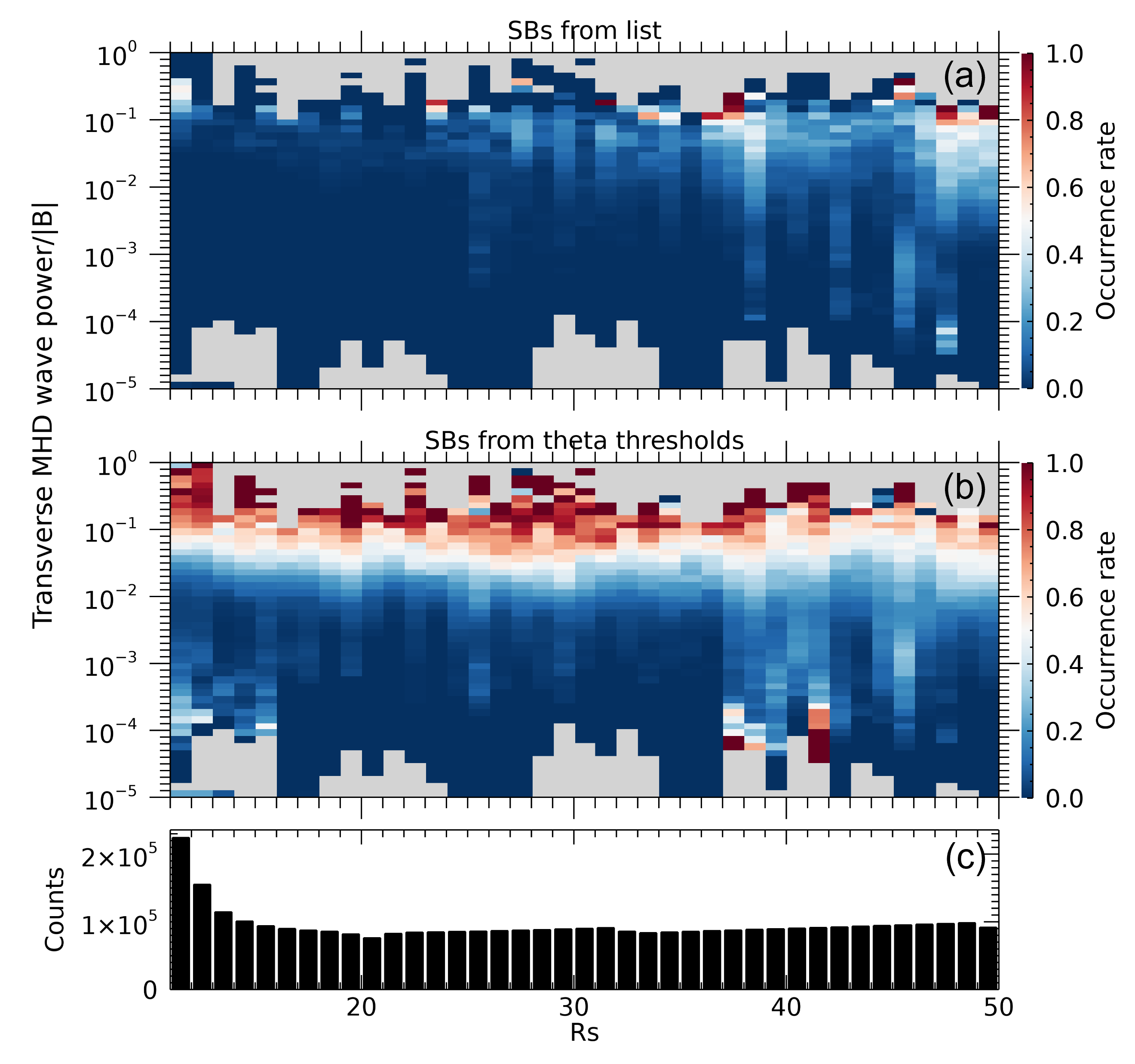}
    \caption{Radial dependence of the occurrence rate of SB intervals as a function of the transverse MHD-band wave power normalized by $|\textbf{B}|$. Panel (a) shows SBs identified from the event list, and panel (b) shows SBs identified from the $\theta$-based threshold method. In each panel, colors indicate the SB occurrence rate in the plane of heliocentric distance \(R_S\) and transverse MHD-band wave power, and gray bins denote regions with no data. Panel (c) shows the radial distribution of the total number of samples used in panels (a) and (b).}
    \label{fig:Bpower_Rs}
\end{figure*}

We next examine how the occurrence of SBs depends on the transverse MHD-band wave power as a function of heliocentric distance, using the $\theta$-based method (Agapitov et al. in prep.) to identify SB intervals across PSP Encounters E17--21. 
\Cref{fig:Bpower_Rs} displays the SB occurrence rate in the plane of heliocentric distance and transverse MHD-band power (normalized by $|\textbf{B}|$), with color indicating the fraction of samples classified as SBs in each two-dimensional bin (relative to all samples in that bin) and gray cells denoting regions with no data. 
Across the full 11-50~$R_S$ range, the color scale reveals that SBs tend to occur more frequently in bins with elevated transverse wave power, with the strongest enhancement seen between about 11 and 35~\(R_S\), while at larger distances the occurrence pattern becomes broader and more diffuse. This trend is present in both \Cref{fig:Bpower_Rs}(a), where SBs are taken from the event-list-based catalog, and \Cref{fig:Bpower_Rs}(b), where they are identified using the $\theta$-threshold method, indicating that enhanced transverse MHD power is a generic feature of SB-rich intervals rather than an artifact of a particular identification scheme.

\subsection{Turbulence nature} 
\label{sec:turbulence}
To further characterize the nature of the enhanced wave power, we examine the spectral index of the magnetic field components as a function of the local field angle $\theta$. 
This analysis tests whether the enhanced wave power observed within SBs is accompanied by changes in the underlying turbulence cascade or primarily reflects an increase in fluctuation amplitude.
The spectral index is evaluated over the 0.01--1~Hz band for both SB and non-SB (NSB) intervals during E01 and E17--21 (Bale et al. 2019; Dudok de Wit et al. 2020). 
This frequency range is chosen to capture the transition from the $1/f$ range to the inertial range, while remaining above the lowest frequencies where spectral estimates become less reliable. We note that, within our E17--21 interval, the $1/f$ range can extend below 0.01 Hz \citep[as explicitly shown for E19 by ][]{huang2025}, which supports using 0.01~Hz as a conservative lower bound rather than the true low-frequency break.

\begin{figure*}
    \centering
    \includegraphics[width=\textwidth]{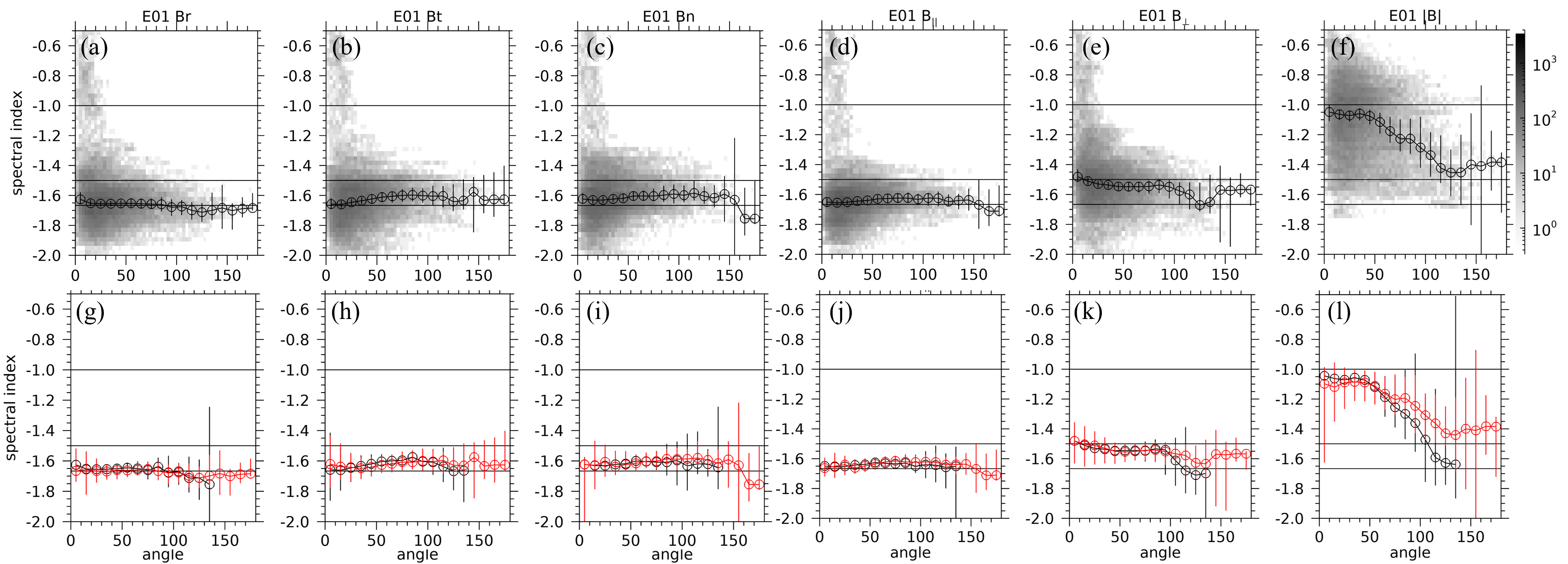}
    \caption{Spectral index of magnetic fluctuations as a function of the local field angle $\theta$ during E01. The upper panels (a-f) show the distribution for non-SB intervals, and the lower panels (g-l) show the corresponding results for SB intervals. Black symbols correspond to non-SB intervals and red symbols to SB intervals. Panels (a-c) and (g-i) present the RTN components ($B_R, B_T, B_N$), while panels (d-f) and (j-l) show the field-aligned components ($B_{\parallel}, B_{\perp}, |\textbf{B}|$). The grayscale background represents the occurrence density, and circles indicate the mean spectral index within each angular bin with error bars denoting the standard deviation. Horizontal lines mark the reference slopes of -1, \(-3/2\), and \(-5/3\).}
    \label{fig:spectral_ind_E1}
\end{figure*}

\Cref{fig:spectral_ind_E1} shows the spectral index as a function of $\theta$ for the RTN, field-aligned components, and magnitude of the magnetic field. 
In all panels, the grayscale background indicates the occurrence density in the $(\theta,\alpha)$ plane, where $\alpha$ is the fitted spectral index, and the circles denote the mean spectral index within each $\theta$ bin with error bars giving the standard deviation.
Horizontal lines mark reference slopes of $-1$, $-3/2$, and $-5/3$.

The RTN components (in Figures \ref{fig:spectral_ind_E1}(a-c,g-i)) and $B_{\parallel}$ (in Figures \ref{fig:spectral_ind_E1}(e,k)) remain close to \(-5/3\) across all angles in both SB and non-SB intervals, indicating that SBs do not substantially alter the underlying inertial-range cascade \citep{bale2019,tenerani2021_evolution}. 
In contrast, $B_{\perp}$ remains near \(\sim -3/2\), consistent with critically balanced Alfvénic turbulence \citep{chen2020}.
Any modest spectral flattening within SB intervals therefore suggests only a marginal enhancement of small-scale transverse power associated with large-scale rotations or intermittency, without substantially affecting the underlying anisotropic cascade.


\begin{figure*}
    \centering
    \includegraphics[width=\textwidth]{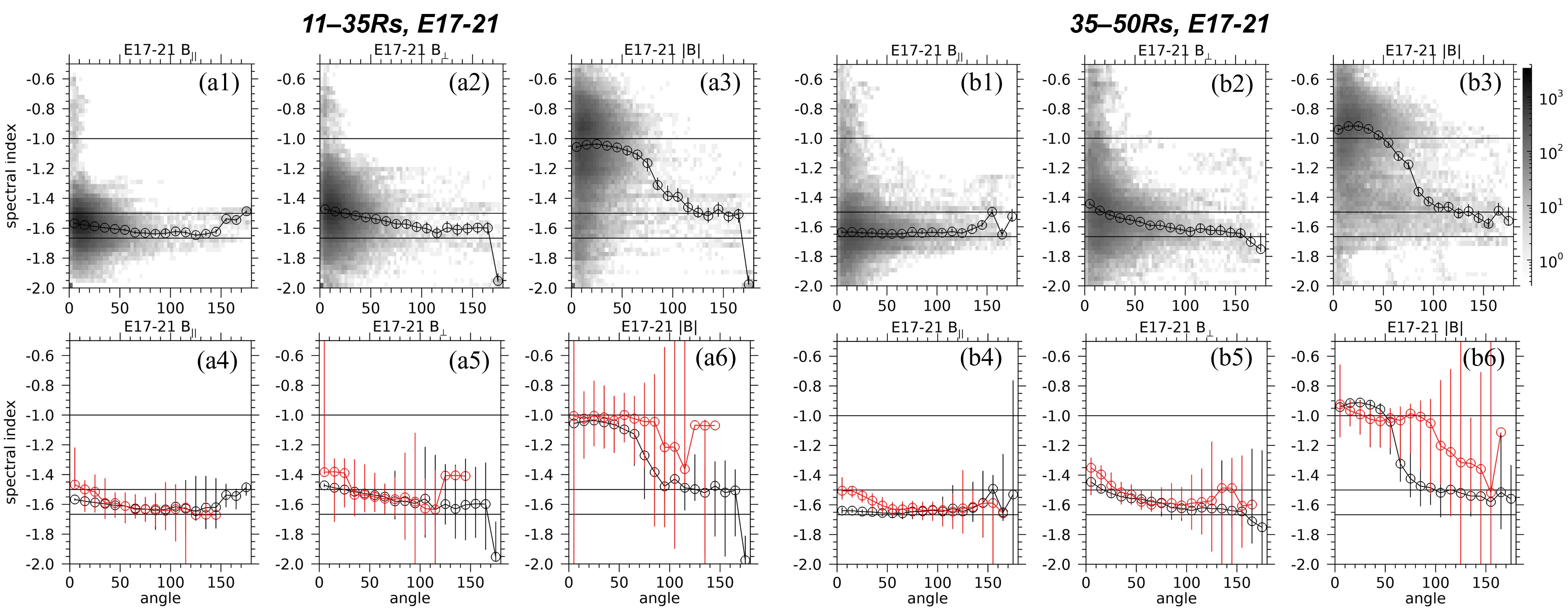}
    \caption{Same format as \Cref{fig:spectral_ind_E1} but for E17--21, showing the spectral index of magnetic fluctuations as a function of the local field angle $\theta$. The left column (panels a1-a6) shows the distributions for 11--35~\(R_S\), and the right column (panels b1-b6) shows the corresponding results for 35--50~\(R_S\). The upper row (a1-b3) presents the field-aligned component $B_{\parallel}$, the transverse component $ B_{\perp}$, and the total magnetic-field magnitude $|\textbf{B}|$, respectively, and the lower row (a4-b6) shows the corresponding results for SB intervals. The grayscale background represents the occurrence density, and circles indicate the mean spectral index within each angular bin with error bars denoting the standard deviation. Horizontal lines mark the reference slopes of -1, \(-3/2\), and \(-5/3\).}
    \label{fig:spectral_ind_E1721}
\end{figure*}

For E17--21, we focus on the field-aligned components, and \Cref{fig:spectral_ind_E1721} shows the spectral index of $B_{\parallel}$, $B_{\perp}$, and $|B|$ as a function of $\theta$ for 11--35~$R_S$ and 35--50~$R_S$. 
The RTN components for E17--21 (not shown) exhibit spectral indices similar to those in E01, clustering around $-5/3$ across angles and SB conditions, and therefore do not provide additional constraints on the turbulence anisotropy.
At both radial ranges, the spectral indices remain consistent with the behavior observed in E01, with $B_{\parallel}$ clustering near $-5/3$ and $B_{\perp}$ near $\sim -3/2$.
Any SB-related differences are limited to slight flattening and increased scatter, reinforcing the conclusion that SBs amplify ion-scale power without fundamentally altering the inertial-range cascade.

To compare how the transition angle depends on heliocentric distance, we examine the $|B|$ spectral index in SB intervals as a function of $\theta$. 
For E01 (35--50~$R_S$), \Cref{fig:spectral_ind_E1} (f), the $|B|$ spectrum remains close to $-1$ at small deflection angles, but steepens toward $\sim -3/2$ as $\theta$ increases. 
At smaller heliocentric distances in E17--21, \Cref{fig:spectral_ind_E1721} (a3,b3), a similar evolution is seen in the 35--50~$R_S$ bin, whereas in the 11--35~$R_S$ bin the mean $|B|$ index approaches $\sim -3/2$ already at intermediate angles. 
These behaviors indicate that the $\theta$-dependence of the $|B|$ index mainly reflects the growing contribution of $B_{\perp}$ to the total-field variance as the fluctuations become more transverse to the mean field and as PSP approaches the Sun. 
Accordingly, the onset angle at which the SB and non-SB $|B|$ spectra begin to separate marks the deflection angle where the enhanced transverse fluctuations become sufficiently important to modify the total-field spectrum.
Furthermore, comparing the onset angle between the SB and non-SB separation in $|B|$ spectra across encounters and radial ranges reveals a systematic trend. 
In E01 (35--50~$R_S$), \Cref{fig:spectral_ind_E1} (l), SB and non-SB intervals begin to diverge only at relatively large deflection angles, with a pronounced difference emerging near $\theta \sim 70^\circ$, whereas in E17--21 the corresponding onset shifts to $\sim 60^\circ$ in the 35--50~$R_S$ bin (\Cref{fig:spectral_ind_E1721} (a6)) and to $\sim 50^\circ$ in the 11--35~$R_S$ bin (\Cref{fig:spectral_ind_E1721} (b6)).
The progressive decrease of this onset angle indicates that progressively smaller magnetic deflections are sufficient for enhanced transverse fluctuations to produce a spectrally distinct $|B|$ signature closer to the Sun.

\section{Discussion and Conclusion} 

\subsection{Geometry versus intrinsic amplification}
We compare magnetic fluctuations in the field-aligned coordinate system to distinguish geometric sampling effects from intrinsic wave amplification within switchbacks.
The results show that SBs amplify sub-proton-scale wave power (0.1-3~\(f_{cp}\), Figures \ref{fig:example}-\ref{fig:Bpower}), while largely preserving the inertial-range spectral scaling (0.01--1~Hz; in Figures \ref{fig:spectral_ind_E1} and \ref{fig:spectral_ind_E1721}), indicating that the enhanced wave activity cannot be explained solely by geometry but also involves intrinsic physical processes, such as shear-driven or reconnection-related energy transfer at SB boundaries, providing observational support for cross-scale energy transfer leading to local wave-driven heating.

While \cite{tatum2024} argued that the enhanced wave power observed during SBs is primarily a consequence of sampling geometry, their analysis was based on integrated magnetic power as a function of the magnetic-field deflection angle relative to the solar-wind flow direction ($\theta_{VB}$) and the SB deflection parameter $z$.
In that interpretation, the apparent power enhancement during SBs is attributed entirely to the spacecraft sampling effect arising from changes in $\theta_{VB}$, which shift the spacecraft from parallel to more perpendicular views of an anisotropic cascade, rather than to intrinsic amplification within the SBs themselves.

In contrast, we compare SB and non-SB intervals at the same local field angle, providing a direct test of whether the enhanced $\delta B_{\perp}$ exceeds the geometric expectation.
Outside SBs, the perpendicular power increases with deflection angle as expected for an anisotropic cascade.
In contrast, SB interiors are systematically shifted toward higher $\delta B_{\perp}$ at the same angles, with the largest excess occurring at small to intermediate deflection angles where geometric projection effects are weakest.

At large deflection angles, the interior and exterior distributions become more similar as projection effects approach saturation, making the small-angle behavior the strongest evidence for intrinsic wave amplification.

Taken together, these results indicate that switchbacks are not simply geometric by-products of sampling an anisotropic cascade, but localized sites of intrinsically enhanced wave activity.
The associated increases in perpendicular magnetic fluctuations, electric-field variability, and proton temperatures support cross-scale transfer of MHD-scale Alfvénic energy into ion-scale fluctuations and heat.

\subsection{Turbulence properties inside switchbacks}

Our spectral-index analysis further constrains how switchbacks interact with the surrounding turbulence. 
The energy distribution of parallel fluctuations appears remarkably universal: $B_{\parallel}$ maintains Kolmogorov-like slopes close to $-5/3$ across angles, radial distances, and SB/non-SB intervals.
The shallower $|B|$ spectra observed in SB intervals, initially suggestive of enhanced compressional fluctuations, are not corroborated by $B_{\parallel}$, which shows no significant difference between SB and non-SB intervals \citep[e.g.,][]{bale2019,dudok_de_wit2020}. 
Instead, the consistent $\sim -3/2$ spectral index of the perpendicular component $B_{\perp}$ across angles and encounters aligns with the shift of SB $|B|$ spectra toward $-3/2$ at large field angles, supporting an interpretation dominated by anisotropic, critically balanced Alfvénic turbulence rather than strong compressional effects \citep[e.g.,][]{iroshnikov1963_IK, kraichnan1965_IK, bruno2019, ervin2025}. 
The slight spectral flattening occasionally seen in SB intervals could be consistent with enhanced small-scale intermittency or additional energy injection at higher frequencies, but our current statistics do not allow us to unambiguously separate such effects from sampling variability; a dedicated intermittency and higher-order moment analysis is left for future work.
Collectively, the tendency for SB $|B|$ spectra to tilt toward a $-3/2$ slope at larger field angles is therefore not evidence for a distinct compressive turbulence regime. 
Instead, it reflects the growing dominance of perpendicular Alfvénic fluctuations in the total-field variance, with this transition occurring from progressively smaller deflection angles as PSP approaches the Sun and the sampling becomes more transverse to the mean field.

\subsection{Implications for heating and cross-scale coupling}

Thus, while switchbacks host enhanced perpendicular wave power at ion-kinetic scales, they do not fundamentally alter the inertial-range spectral slopes, which remain consistent with anisotropic, critically balanced Alfvénic turbulence in both switchback and non-switchback intervals.

In summary, we find that
\begin{enumerate}
\item perpendicular magnetic fluctuations at ion-kinetic scales are systematically stronger inside switchbacks than in neighboring solar-wind intervals, even when compared at the same local field angle;
\item the enhancement of $\delta B_{\perp}$ inside switchbacks is most pronounced at small and intermediate deflection angles, where simple geometric projection of an anisotropic cascade would normally lead to relatively weak power, implying that geometry alone cannot fully account for the observed wave energy; and
\item intervals of enhanced perpendicular power coincide with proton heating and inertial-range spectra that remain consistent with anisotropic, critically balanced Alfvénic turbulence, supporting a scenario in which switchbacks tap the ambient cascade to drive localized ion-scale dissipation.
\end{enumerate}

These findings suggest that magnetic switchbacks are not merely passive signatures of large-scale Alfvénic structures but active elements of near-Sun solar-wind turbulence that shape how energy is redistributed and dissipated between scales and particle populations \citep{mozer2026}.

\section*{Acknowledgments}
\begin{acknowledgments}
KEC, OVA, and LC were supported by NASA contracts  80NSSC22K0433, 80NNSC19K0848, 80NSSC21K1770, and NASA’s Living with a Star (LWS) program (contract 80NSSC20K0218). 

We thank the NASA Parker Solar Probe Mission, the SWEAP team led by Justin Kasper, and the FIELDS team led by Stuart Bale for the use of the data.
Parker Solar Probe was designed and built and is now operated by the Johns Hopkins Applied Physics Laboratory as part of NASA’s Living with a Star (LWS) program (contract NNN06AA01C)

\end{acknowledgments}

\bibliography{ref}{}
\bibliographystyle{aasjournalv7}

\end{document}